\documentclass[prl,superscriptaddress,reprint]{revtex4-1} 

\usepackage{graphics}
\usepackage{hyperref}
\newcommand{\ket}[1]{\ensuremath{\left|#1\right>}}%

\newcommand{\figref}[2]{Fig. \ref{#1}{\bf{}#2}}%
\newcommand{\dbz}{\ensuremath{\Delta{}B_{z}}}%

\begin{document}

\title{Suppressing qubit dephasing using real-time Hamiltonian estimation}
 \date{\today}
\author{M. D. Shulman}
\author{S. P. Harvey}
\author{J. M. Nichol}
\affiliation{Department of Physics, Harvard University, Cambridge, MA,
  02138, USA}

\author{S. D. Bartlett}
\author{A. C. Doherty}
\affiliation{Centre for Engineered Quantum Systems, School of Physics,
  The University of Sydney, Sydney, NSW 2006, Australia}

\author{V. Umansky}
\affiliation{Braun Center for Submicron Research, Department of Condensed
  Matter Physics, Weizmann Institute of Science, Rehovot 76100 Israel}

\author{A. Yacoby}
\affiliation{Department of Physics, Harvard University, Cambridge, MA,
  02138, USA}
\email[email:]{yacoby@physics.harvard.edu}

\begin{abstract}

Unwanted interaction between a quantum system and its fluctuating
environment leads to decoherence and is the primary obstacle to
establishing a scalable quantum information processing
architecture. Strategies such as environmental
\cite{Hite2012,Bluhm2010} and materials engineering \cite{Paik2011},
quantum error correction \cite{Nielsen2000,Reed2011} and dynamical
decoupling \cite{Hahn1950} can mitigate decoherence, but generally
increase experimental complexity. Here we improve coherence in a qubit
using real-time Hamiltonian parameter estimation
\cite{Wiseman2010}. Using a rapidly converging Bayesian approach, we
precisely measure the splitting in a singlet-triplet spin qubit faster
than the surrounding nuclear bath fluctuates. We continuously adjust
qubit control parameters based on this information, thereby improving
the inhomogenously broadened coherence time ($T_{2}^{*}$) from tens of
nanoseconds to above 2 $\mu$s and demonstrating the effectiveness of
Hamiltonian estimation in reducing the effects of correlated noise in
quantum systems. Because the technique demonstrated here is compatible
with arbitrary qubit operations, it is a natural complement to quantum
error correction and can be used to improve the performance of a wide
variety of qubits in both metrological and
quantum-information-processing applications.
\end{abstract}

\maketitle

Hamiltonian parameter estimation is a rich field of active
experimental and theoretical research that enables precise
characterization and control of quantum systems
\cite{Wiseman2010}. For example, magnetometry schemes employing
Hamiltonian learning have demonstrated dynamic range and sensitivities
exceeding those of standard methods
\cite{Waldherr2012,Nusran2012}. Such applications focused on
estimating parameters that are quasistatic on experimental
timescales. However, the effectiveness of Hamiltonian learning also
offers exciting prospects for estimating fluctuating parameters
responsible for decoherence in quantum systems. In this work we employ
techniques from Hamiltonian estimation to prolong the coherence of a
qubit by more than a factor of 30. Importantly, our estimation
protocol, which is based on recent theoretical work
\cite{Sergeevich2011}, requires relatively few measurements
($\approx$100) which we perform rapidly enough (total time
$\approx\mathrm{100}\mu{}s$) to resolve the qubit splitting faster
than its characteristic fluctuation time. We adopt a paradigm in which
we separate experiments into ``estimation'' and ``operation''
segments, and we use information from the former to optimize control
parameters for the latter in real-time. Our method dramatically
prolongs coherence without using complex pulse sequences such as those
required for non-identity dynamically decoupled operations
\cite{Kestner2013}.

\begin{figure}
\resizebox{\columnwidth}{!}
{\includegraphics{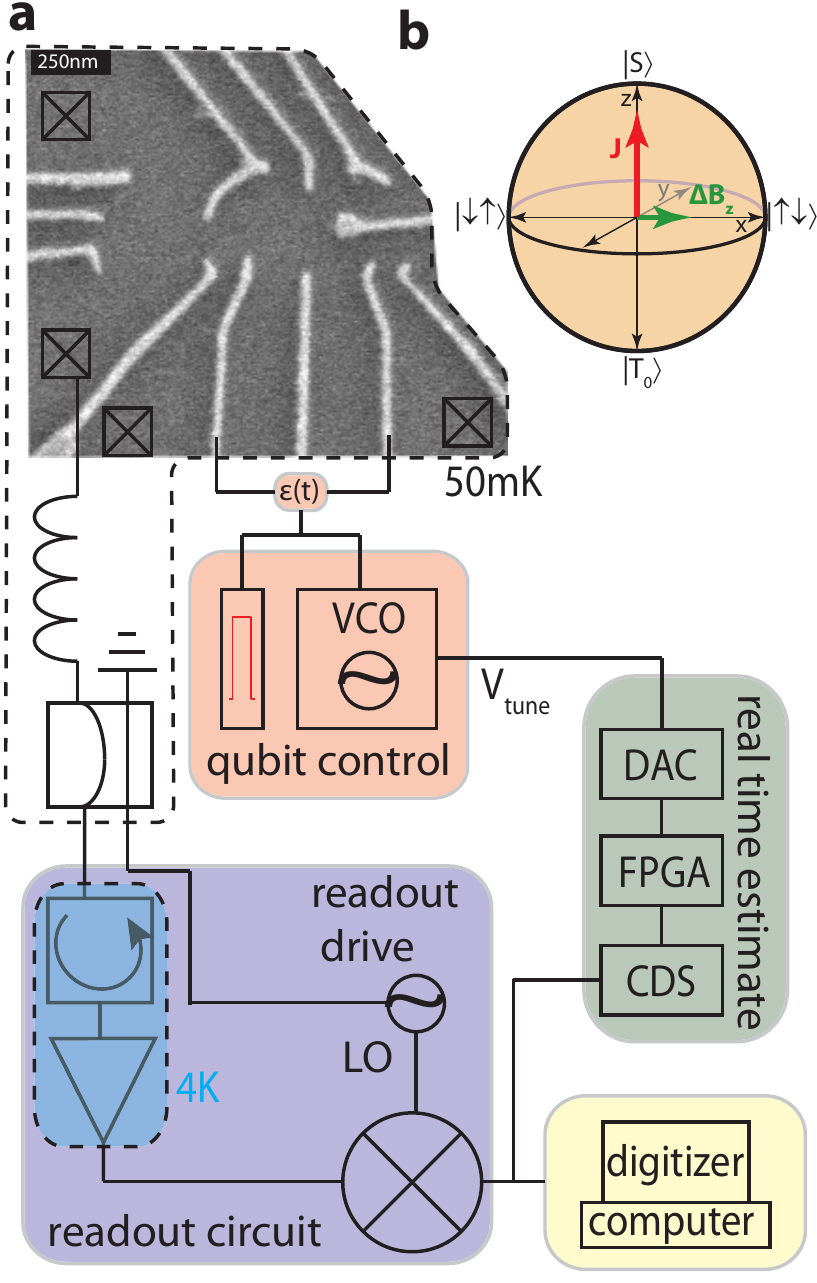}}
\caption{ Experimental apparatus. \textbf{a,} A scanning electron
  microscope image of the double QD with a schematic of the apparatus
  used for adaptive qubit control. A floating metal gate protruding
  from the right can be seen which increases the capacitance between
  the qubit and an adjacent qubit (not pictured), which is left
  inactive for this work. The reflected readout drive signal is
  demodulated to DC, digitized by a correlated double sampler (CDS),
  and \dbz{} is estimated in real time by the field programmable gate
  array (FPGA). The FPGA updates the digital to analog converter (DAC)
  in order to keep the voltage controlled oscillator (VCO) resonant
  with the estimated value of \dbz{}. The VCO controls the voltage
  detuning, $\epsilon(t)$ between the QDs, which, in turn, modulates
  $J$ at $\Omega_{J}$. \textbf{b,} The Bloch sphere representation for
  the $S$-$T_{0}$ qubit showing the two axes of control, $J$ and \dbz.
\label{device}
}
\end{figure}

The singlet-triplet ($S$-$T_0$) qubit ~\cite{Petta2005,Maune2011}
studied in this work is formed by two gate-defined lateral quantum
dots (QDs) in a GaAs/AlGaAs heterostructure (\figref{device}{a})
similar to that of refs.~\cite{Dial2013,Shulman2012}. The qubit can be
rapidly initialized in the singlet state \ket{S} in $\approx 20$ ns
and read out with $98 \%$ fidelity in $\approx 1$ $
\mu{}s$~\cite{Barthel2009,Reilly2007} (see SOM). Universal quantum
control is provided by two distinct drives~\cite{Foletti2009}: the
exchange splitting, $J$, between \ket{S} and \ket{T_0}, and the
magnetic field gradient, \dbz, due to the hyperfine interaction with
host Ga and As nuclei. The Bloch sphere representation for this qubit
can be seen in \figref{device}{b}.

\begin{figure*}
\resizebox{\textwidth}{!}{\includegraphics{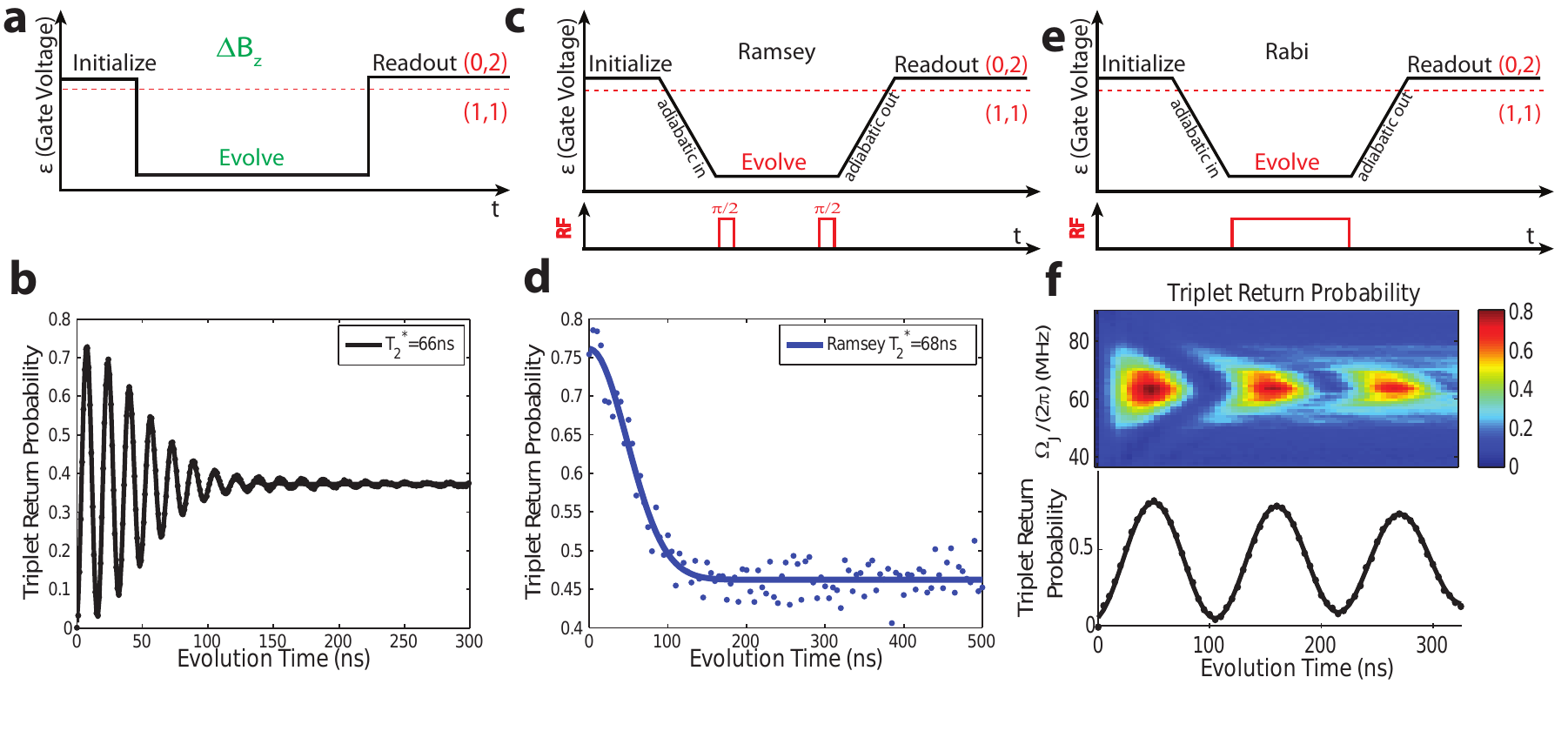}}
\caption{ \dbz{} oscillations. \textbf{a,} The pulse sequence used
  to estimate \dbz{}. \textbf{b,}. Using nuclear feedback, \dbz{}
  oscillations decay in a coherence time
  $T_{2}^{*}\approx{}\mathrm{70ns}$ due to residual slow fluctuations
  in \dbz{}. \textbf{c,} The Ramsey sequence used to operate the
  $S$-$T_{0}$ qubit in the rotating frame. \textbf{d,} The Ramsey
  contrast decays similarly to the oscillations in \textbf{(c)} due to
  the same residual slow fluctuations in \dbz{}. \textbf{e,} The
  Rabi pulse sequence used to drive the qubit in the rotating
  frame. \textbf{f,} The rotating frame $S$-$T_0$ qubit exhibits the
  typical behavior when sweeping drive frequency and time (top). When
  driven on resonance (bottom), the qubit undergoes Rabi oscillations,
  demonstrating control in the rotating frame.
\label{dbz}
}
\end{figure*}

In this work, we focus on qubit evolution around \dbz{}
(\figref{dbz}{a}). Due to statistical fluctuations of the nuclei,
\dbz{} varies randomly in time, and consequently oscillations around
this field gradient decay in a time $T_2^* \approx$ 10
ns~\cite{Petta2005}. A nuclear feedback scheme relying on dynamic
nuclear polarization~\cite{Bluhm2010} can be employed to set the mean
gradient, ($g^* \mu_B \dbz /h \approx 60$ MHz in this work) as well as
reduce the variance of the fluctuations. Here, $g^* \approx -0.44$ is
the effective gyromagnetic ratio in GaAs, $\mu_B$ is the Bohr magneton
and $h$ is Planck's constant. In what follows, we adopt units where
$g^* \mu_B /h = 1$.  With the use of this feedback, the coherence time
improves to $T_2^* \approx 100$ ns ~\cite{Bluhm2010}
(\figref{dbz}{b}), limited by the low nuclear pumping efficiency
\cite{Foletti2009}. Crucially, the residual fluctuations are
considerably slower than the timescale of qubit operations
\cite{Bluhm2011}. To take advantage of these slow dynamics, we
introduce a method that measures the fluctuations and manipulates the
qubit based on precise knowledge but not precise control of the
environment.

\begin{figure}
\resizebox{\columnwidth}{!}{\includegraphics{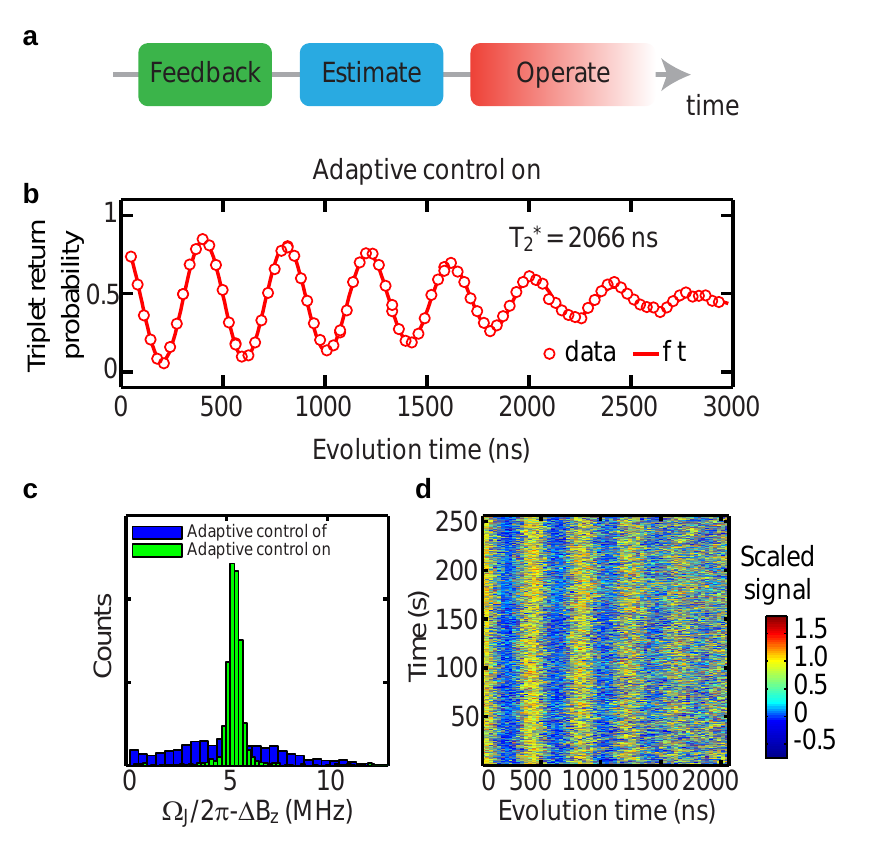}}
 \caption{ Adaptive control. \textbf{a,} For these measurements we
   first perform our standard nuclear feedback, then quickly estimate
   \dbz{} and update the qubit control, then operate the qubit at the
   correct driving frequency. \textbf{b,} Using adaptive control, we
   perform a Ramsey experiment (deliberately detuned to see
   oscillations) and obtain coherence times larger than $2$
   $\mu$s. \textbf{c,} Histograms of measured Ramsey detunings with
   and without adaptive control. For clarity, these data were taken
   with a different mean detuning than those in
   \textbf{(b)}. \textbf{d,} Raw data for 1024 consecutive Ramsey
   experiments with adaptive control lasting 250 s in total. A value
   of $1$ corresponds to \ket{T_0} and $0$ corresponds to
   \ket{S}. Stabilized oscillations are clearly visible in the data,
   demonstrating the effect of adaptive control.
\label{FPGA}
}
\end{figure}

We operate the qubit in the rotating frame of \dbz, where qubit
rotations are driven by modulating $J$ at the frequency
$\frac{\Omega_{J}}{2\pi}=\dbz{}$ \cite{Coish2004,Klauser2006}. To
measure Rabi oscillations, the qubit is adiabatically prepared in the
ground state of \dbz{} (\ket{\psi}=\ket{\uparrow{}\downarrow}), and an
oscillating $J$ is switched on (\figref{dbz}{e}), causing the qubit to
precess around $J$ in the rotating frame. Additionally, we perform a
Ramsey experiment (\figref{dbz}{c}) to determine $T_{2}^{*}$, and as
expected, we observe the same decay (\figref{dbz}{d}) as
\figref{dbz}{b}. More precisely, the data in \figref{dbz}{d} represent
the average of 1024 experimental repetitions of the same qubit
operation sequence immediately following nuclear feedback. The
feedback cycle resets \dbz{} to its mean value (60MHz) with residual
fluctuations of $(\sqrt{2}\pi{}T_{2}^{*})^{-1}\approx{}10\mathrm{MHz}$
between experimental repetitions. However, within a given experimental
repetition, \dbz{} is approximately constant.  Therefore we present an
adaptive control scheme where, in following nuclear feedback, we
quickly estimate \dbz{} and tune $\frac{\Omega_{J}}{2\pi}=\dbz{}$ in
order to prolong qubit coherence (\figref{FPGA}{a}).

To estimate \dbz{}, we repeatedly perform a series of singleshot
measurements after allowing it to evolve around \dbz{} for some amount
of time (\figref{dbz}{a}). Rather than fixing this evolution time to
be constant for all trials, we make use of recent theoretical results
in Hamiltonian parameter estimation
~\cite{Sergeevich2011,Ferrie2013,Klauser2006} and choose linearly
increasing evolution times, $t_k = k t_{samp}$, where
$k=1,2,\cdots,N$. We choose the sampling time $t_{samp}$ such that the
estimation bandwidth $\mathcal{B} =\frac{1}{2t_{samp}}$ is several
times larger than the magnitude of the residual fluctuations in \dbz,
roughly 10 MHz. With a Bayesian approach to estimate \dbz{} in
real-time, the longer evolution times (large $k$) leverage the
increased precision obtained from earlier measurements to provide
improved sensitivity, allowing the estimate to outperform the standard
limit associated with repeating measurements at a single evolution
time. Denoting the outcome of the $k^{\mathrm{th}}$ measurement as
$m_{k}$ (either \ket{S} or \ket{T_0}), we define $P(m_k|\dbz)$ as the
conditional probability for $m_{k}$ given a value \dbz{}.  We write
\begin{equation}
P(m_k|\dbz{})=\frac{1}{2}\left[1+r_{k}\left(\alpha{}+\beta{}\cos\left(2\pi{}\dbz{}t_{k}\right)\right)\right],\label{model}
\end{equation}
where $r_{k}$=$1$ $(-1)$ for $m_{k}$=\ket{S}
$\left(\ket{T_{0}}\right)$, and $\alpha=0.25$ and $\beta=0.67$ are
parameters determined by the measurement error and axis of rotation on
the Bloch sphere (see SOM). Since measurement outcomes are assumed to
be independent, we write the conditional probability for \dbz{} given
the results of $N$ measurements as:
\begin{eqnarray}
P(\dbz{}|m_{N},m_{N-1},...m_{1})\\
=P(\dbz{}|m_{N-1},...m_{1})\cdot{}P(\dbz{}|m_{N})\\
= \prod_{k=1}^{N}P(\dbz|m_k).\label{posterior}
\end{eqnarray}
Using Bayes' rule, i.e., $P(\dbz|m_k)=P(m_k|\dbz)P(\dbz)/P(m_k)$, and
eq.~\ref{model}, we can rewrite eq. ~\ref{posterior} as:
\begin{eqnarray}
 \nonumber P(\dbz{}|m_{N},m_{N-1},...m_{1})= \\
P_{0}(\dbz{})\mathcal{N}\prod_{k=1}^{N}\left(1+r_{k}\left(\alpha{}+\beta{}
  \cos \left(2\pi{}\dbz{}t_{k}\right)\right)\right],
\end{eqnarray}
where $\mathcal{N}$ is a normalization constant and $P_{0}(\dbz{})$ is
a prior distribution to which the algorithm is empirically
insensitive, and which we take to be a constant over the estimation
bandwidth. After the last measurement, we find the value of \dbz~ that
maximizes the posterior distribution
$P(\dbz{}|m_{N},m_{N-1},...m_{1})$.

We implement this algorithm in real time on a field-programmable gate
array (FPGA) for 256 linearly spaced frequencies between 50 and
70MHz. With each measurement $m_k$, the readout signal is digitized
and passed to the FPGA, which computes the Bayesian estimate of \dbz{}
and updates an analog voltage that tunes the frequency of a voltage
controlled oscillator (\figref{device}{a}) (see SOM). Following the
$N^{\mathrm{th}}$ sample, $\frac{\Omega_{J}}{2\pi}$ nearly matches
\dbz{}, and since the nuclear dynamics are slow, the qubit can be
operated with long coherence without any additional complexity. To
quantify how well the FPGA estimate matches \dbz{}, we perform a
Ramsey experiment (deliberately detuned to observe oscillations) with
this real-time tracking of \dbz{} and find optimal performance for
$N\approx{}120$, with a maximum experimental repetition rate, limited
by the FPGA, of 250kHz and a sampling time $t_{samp}=12$ ns. Under
these conditions, we observe $T_{2}^{*}=2066$ ns, a 30-fold increase
in coherence (\figref{FPGA}{b}). We note that these data are taken
with the same pulse sequence as those in \figref{dbz}{d}. To further
compare qubit operations with and without this technique, we measure
Ramsey fringes for $\approx$ 250s (\figref{FPGA}{d}), and histogram
the observed Ramsey detunings. With adaptive control we observe a
stark narrowing of the observed frequency distribution, consistent
with this improved coherence (\figref{FPGA}{c}).

\begin{figure}
\resizebox{\columnwidth}{!}{\includegraphics{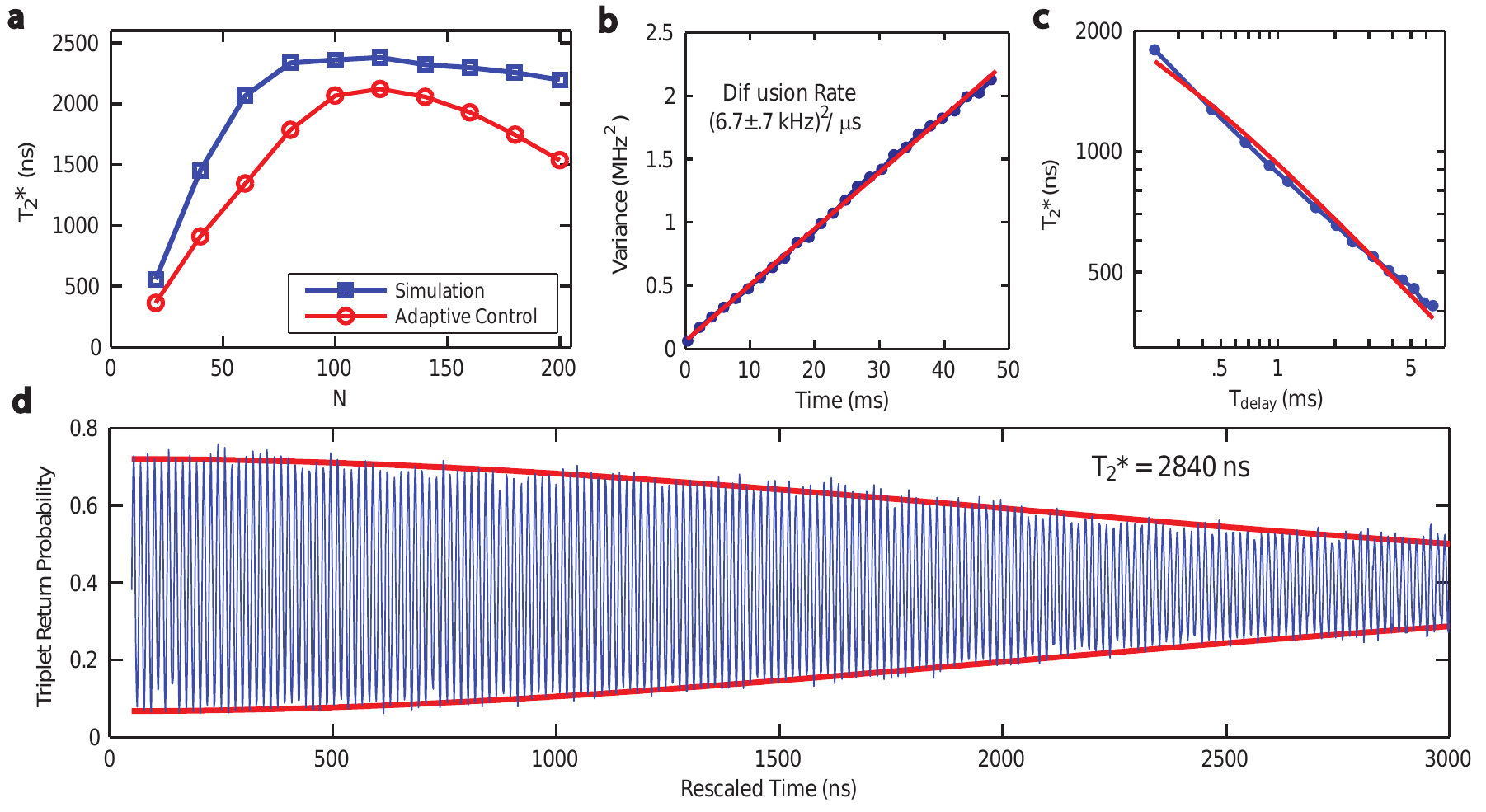}}
\caption{ \dbz{} diffusion. \textbf{a,} The coherence time,
  $T_{2}^{*}$ using the adaptive control and for a simulation show a
  peak, indicating that there is an optimal number of measurements to
  make when estimating \dbz{}. \textbf{b,} When many time traces of
  \dbz{} are considered, their variance grows linearly with time,
  indicating a diffusion process. \textbf{c,} The scaling of
  $T_{2}^{*}$ as a function of $T_{delay}$ for software scaled data is
  consistent with diffusion of \dbz{}. \textbf{d,} The performance
  of the Bayesian estimate of \dbz{} can be estimated using software
  post processing, giving $T_{2}^{*}=\mathrm{2840ns}$, which
  corresponds to a precision of $\sigma_{\dbz{}}=\mathrm{80kHz}$.
\label{software}
}
\end{figure}

Although the estimation scheme employed here is theoretically
predicted to improve monotonically with $N$ \cite{Sergeevich2011}, we
find that there is an optimum ($N\approx{}120$), after which $T_2^*$
slowly decreases with increasing $N$ (\figref{software}{a}). A
possible explanation for this trend is fluctuation of the nuclear
gradient during the estimation period. To investigate this, we obtain
time records of \dbz{} using the Bayesian estimate and find that its
variance increases linearly in time at the rate of $( \mathrm{6.7}
\pm.7 \mathrm{kHz})^{2}\mu{}s^{-1}$ (\figref{software}{c}). The
observed linear behavior suggests a model where the nuclear gradient
diffuses, and using the measured diffusion of \dbz{}, we simulate the
performance of the Bayesian estimate as a function of $N$(see
SOM). Given that the simulation has no free parameters, we find good
agreement with the observed $T_{2}^{*}$, indicating that indeed,
diffusion limits the accuracy with which we can measure \dbz{}
(\figref{software}{a}).

This model suggests that increasing the rate of measurements during
estimation will improve the accuracy of the Bayesian estimate. Because
our FPGA limits the repetition rate of qubit operations to 250 kHz, we
demonstrate the effect of faster measurements through software
post-processing with the same Bayesian estimate. To do so, we first
use the same estimation sequence, but for the operation segment, we
measure the outcome after evolving around \dbz{} for a single
evolution time, $t_{evo}$, rather than performing a rotating frame
Ramsey experiment, and we repeat this experiment a total of $N_{tot}$
times. In processing, we perform the Bayesian estimate of each
$\dbz{}_{,i}$, sort the data by adjusted time
$\tau_{i}=\frac{\dbz{}_{,i}t_{evo,i}}{\langle\dbz{}\rangle}$ (for
$i=1,2,\cdots, N_{tot}$), and average together points of similar
$\tau$ in order to observe oscillations (see SOM). We fit the decay of
these oscillations to extract $T_{2}^{*}$ and the precision of the
Bayesian estimate,
$\sigma_{\dbz}=\left(\sqrt{2}\pi{}T_{2}^{*}\right)^{-1}$ (see
SOM). For the same operation and estimation parameters, we find that
$T_{2}^{*}$ extracted from software post-processing agrees with that
extracted from adaptive control (see SOM). Using a repetition rate as
high as 667kHz, we show coherence times above 2800 ns, corresponding
to an error of $\sigma_{\dbz{}}$=80kHz (\figref{software}{d}),
indicating that improvements are easily attainable by using faster
(commercially available) FPGAs.

Additionally, we use this post-processing technique to examine the
effect of this technique on the duty cycle of experiments as well as
the stability of the \dbz{} estimate. To do so we introduce a delay
$T_{delay}$ between the estimation of \dbz{} and the single evolution
measurement performed in place of the operation. We find $T_{2}^{*} =
(a+bT_{delay}^{c})^{-0.5}$, where c = 0.99 (\figref{software}{c}),
consistent with diffusion of \dbz{}. Indeed, this dependence
underscores the potential of adaptive control, since it demonstrates
that after a single estimation sequence, the qubit can be operated for
$>\mathrm{1}$ms with $T_{2}^{*}>\mathrm{1}\mu{}\mathrm{s}$. Thus,
adaptive control need not significantly reduce the experimental duty
cycle.

In this work, we have used real-time adaptive control based on
Hamiltonian parameter estimation of a $S-T_0$ spin qubit to prolong
$T_{2}^{*}$ from 70ns to more than 2 $\mu{}s$. Dephasing due to
nuclear spins has long been considered a significant obstacle to
quantum information processing using semiconductor spin
qubits~\cite{Schroer2008}.  However, here we have shown that with a
combination of nuclear feedback and real-time Hamiltonian estimation,
we are able to achieve ratios of coherence times to operation times in
excess of 200 without recourse to dynamical decoupling. If the same
adaptive control techniques were applied to gradients as high as 1
GHz~\cite{Foletti2009}, ratios exceeding 4000 would be possible, and
longer coherence times may be attainable with more sophisticated
techniques~\cite{Sergeevich2011}.  The method we have presented is
straightforward to implement, compatible with arbitrary qubit
operations, and general to all qubits that suffer from non-Markovian
noise.  Looking ahead, it is likely, therefore, to play a key role in
realistic quantum error correction efforts, where even modest
improvements in baseline error rate greatly diminish experimental
complexity and enhance prospects for a scalable quantum information
processing architecture.

\begin{acknowledgements}

M.D.S, S.P.H, and J.M.N. contributed equally to this
work. Correspondence should be directed to AY at
yacoby@physics.harvard.edu We acknowledge James MacArthur for building
the correlated double sampler. This research was funded by the United
States Department of Defense, the Office of the Director of National
Intelligence (ODNI), Intelligence Advanced Research Projects Activity
(IARPA), and the Army Research Office grant (W911NF-11-1-0068 and
W911NF-11-1-0068). All statements of fact, opinion or conclusions
contained herein are those of the authors and should not be construed
as representing the official views or policies either expressly or
implied of of IARPA, the ODNI, or the U.S. Government. S.P.H was
supported by the Department of Defense (DoD) through the National
Defense Science \& Engineering Graduate Fellowship (NDSEG)
Program. ACD acknowledges discussions with Matthew Wadrop regarding
extracting diffusion constants. ACD and SDB acknowledge support from
the ARC via the Centre of Excellence in Engineering Quantum Systems
(EQuS) project number CE110001013. This work was performed in part at
the Center for Nanoscale Systems (CNS), a member of the National
Nanotechnology Infrastructure Network (NNIN), which is supported by
the National Science Foundation under NSF award no. ECS-0335765. CNS
is part of Harvard University.
\end{acknowledgements}

\bibliography{../../bibtex_2012_04_04}{} \bibliographystyle{unsrtnat}

\begin{thebibliography}{23}
\providecommand{\natexlab}[1]{#1}
\providecommand{\url}[1]{\texttt{#1}}
\expandafter\ifx\csname urlstyle\endcsname\relax
  \providecommand{\doi}[1]{doi: #1}\else
  \providecommand{\doi}{doi: \begingroup \urlstyle{rm}\Url}\fi

\bibitem[Hite et~al.(2012)Hite, Colombe, Wilson, Brown, Warring, J\"ordens,
  Jost, McKay, Pappas, Leibfried, and Wineland]{Hite2012}
D.~A. Hite, Y.~Colombe, A.~C. Wilson, K.~R. Brown, U.~Warring, R.~J\"ordens,
  J.~D. Jost, K.~S. McKay, D.~P. Pappas, D.~Leibfried, and D.~J. Wineland.
\newblock 100-fold reduction of electric-field noise in an ion trap cleaned
  with \emph{in situ} argon-ion-beam bombardment.
\newblock \emph{Phys. Rev. Lett.}, 109:\penalty0 103001, Sep 2012.
\newblock \doi{10.1103/PhysRevLett.109.103001}.
\newblock URL \url{http://link.aps.org/doi/10.1103/PhysRevLett.109.103001}.

\bibitem[Bluhm et~al.(2010)Bluhm, Foletti, Mahalu, Umansky, and
  Yacoby]{Bluhm2010}
H.~Bluhm, S.~Foletti, D.~Mahalu, V.~Umansky, and A.~Yacoby.
\newblock Enhancing the coherence of a spin qubit by operating it as a feedback
  loop that controls its nuclear spin bath.
\newblock \emph{Phys. Rev. Lett.}, 105:\penalty0 216803, 2010.

\bibitem[Paik et~al.(2011)Paik, Schuster, Bishop, Kirchmair, Catelani, Sears,
  Johnson, Reagor, Frunzio, Glazman, Girvin, Devoret, and Schoelkopf]{Paik2011}
Hanhee Paik, D.~I. Schuster, Lev~S. Bishop, G.~Kirchmair, G.~Catelani, A.~P.
  Sears, B.~R. Johnson, M.~J. Reagor, L.~Frunzio, L.~I. Glazman, S.~M. Girvin,
  M.~H. Devoret, and R.~J. Schoelkopf.
\newblock Observation of high coherence in {J}osephson junction qubits measured
  in a three-dimensional circuit {QED} architecture.
\newblock \emph{Phys. Rev. Lett.}, 107:\penalty0 240501, Dec 2011.
\newblock \doi{10.1103/PhysRevLett.107.240501}.
\newblock URL \url{http://link.aps.org/doi/10.1103/PhysRevLett.107.240501}.

\bibitem[Nielsen and Chuang(2000)]{Nielsen2000}
M.A. Nielsen and I.L. Chuang.
\newblock \emph{Quantum Computation and Quantum Information}.
\newblock Cambridge University Press, 2000.

\bibitem[Reed et~al.(2011)Reed, DiCarlo, Nigg, Sun, Frunzio, Girvin, and
  Schoelkopf]{Reed2011}
M.D. Reed, L.~DiCarlo, S.E. Nigg, L.~Sun, L~Frunzio, S.~M. Girvin, and R.~J.
  Schoelkopf.
\newblock Realization of a three-qubit quantum error correction with
  superconducting circuits.
\newblock \emph{Nature}, 482:\penalty0 382, 2011.

\bibitem[Hahn(1950)]{Hahn1950}
E.~L. Hahn.
\newblock Spin echoes.
\newblock \emph{Phys. Rev.}, 80:\penalty0 580, 1950.

\bibitem[Wiseman and Milburn(2010)]{Wiseman2010}
Howard~M. Wiseman and Gerard~J. Milburn.
\newblock \emph{Quantum Measurement and Control}.
\newblock Cambridge University Press, 2010.

\bibitem[Waldherr et~al.(2012)Waldherr, Beck, Neumann, Said, Nitsche, Markham,
  Twitchen, Twamley, Jelezko, and Wrachtrup]{Waldherr2012}
G.~Waldherr, J.~Beck, P.~Neumann, R.~S. Said, M.~Nitsche, M.~L. Markham, D.~J.
  Twitchen, J.~Twamley, F.~Jelezko, and J.~Wrachtrup.
\newblock {High-dynamic-range magnetometry with a single nuclear spin in
  diamon}.
\newblock \emph{Nature Nanotechnology}, 7:\penalty0 105--108, 2012.
\newblock \doi{10.1038/nnano.2011.224}.

\bibitem[Nusran et~al.(2012)Nusran, Ummal, and Dutt]{Nusran2012}
N.~M. Nusran, M.~Ummal, and M.V~Gurudev Dutt.
\newblock High dynamic range magnetometry with a single electronic spin in
  diamon.
\newblock \emph{Nature Nanotechnology}, 7:\penalty0 109, 2012.
\newblock \doi{10.1038/nnano.2011.225}.

\bibitem[Sergeevich et~al.(2011)Sergeevich, Chandran, Combes, Bartlett, and
  Wiseman]{Sergeevich2011}
Alexandr Sergeevich, Anushya Chandran, Joshua Combes, Stephen~D. Bartlett, and
  Howard~M. Wiseman.
\newblock Characterization of a qubit hamiltonian using adaptive measurements
  in a fixed basis.
\newblock \emph{Phys. Rev. A}, 84:\penalty0 052315, Nov 2011.
\newblock \doi{10.1103/PhysRevA.84.052315}.
\newblock URL \url{http://link.aps.org/doi/10.1103/PhysRevA.84.052315}.

\bibitem[Kestner et~al.(2013)Kestner, Wang, Bishop, Barnes, and
  Das~Sarma]{Kestner2013}
J.~P. Kestner, Xin Wang, Lev~S. Bishop, Edwin Barnes, and S.~Das~Sarma.
\newblock Noise-resistant control for a spin qubit array.
\newblock \emph{Phys. Rev. Lett.}, 110:\penalty0 140502, Apr 2013.
\newblock \doi{10.1103/PhysRevLett.110.140502}.
\newblock URL \url{http://link.aps.org/doi/10.1103/PhysRevLett.110.140502}.

\bibitem[Petta et~al.(2005)Petta, Johnson, Taylor, Laird, Yacoby, Lukin,
  Marcus, Hanson, and Gossard]{Petta2005}
J.~R. Petta, A.~C. Johnson, J.~M. Taylor, E.~A. Laird, A.~Yacoby, M.~D. Lukin,
  C.~M. Marcus, M.~P. Hanson, and A.~C. Gossard.
\newblock Coherent manipulation of coupled electron spins in semiconductor
  quantum dots.
\newblock \emph{Science}, 309:\penalty0 2180, 2005.

\bibitem[Maune et~al.(2011)Maune, Borselli, Huang, Ladd, Deelman, Holabird,
  Kiselev, Alvarado-Rodriguez, Ross, Schmitz, Sokolich, Watson, Gyure, and
  Hunter]{Maune2011}
B.~M. Maune, M.~G. Borselli, B.~Huang, T.~D. Ladd, P.~W. Deelman, K.~S.
  Holabird, A.~A. Kiselev, I.~Alvarado-Rodriguez, R.~S. Ross, A.~E. Schmitz,
  M.~Sokolich, C.~A. Watson, M.~F. Gyure, and A.~T. Hunter.
\newblock Coherent singlet-triplet oscillations in a silicon-based double
  quantum dot.
\newblock \emph{Nature}, 481:\penalty0 344--347, 2011.
\newblock \doi{10.1038/nature10707}.

\bibitem[Dial et~al.(2013)Dial, Shulman, Harvey, Bluhm, Umansky, and
  Yacoby]{Dial2013}
O.~E. Dial, M.~D. Shulman, S.~P. Harvey, H.~Bluhm, V.~Umansky, and A.~Yacoby.
\newblock Charge noise spectroscopy using coherent exchange oscillations in a
  singlet-triplet qubit.
\newblock \emph{Phys. Rev. Lett.}, 110:\penalty0 146804, Apr 2013.
\newblock \doi{10.1103/PhysRevLett.110.146804}.
\newblock URL \url{http://link.aps.org/doi/10.1103/PhysRevLett.110.146804}.

\bibitem[Shulman et~al.(2012)Shulman, Dial, Harvey, Bluhm, Umansky, and
  Yacoby]{Shulman2012}
M.~D. Shulman, O.~E. Dial, S.~P. Harvey, H.~Bluhm, V.~Umansky, and A.~Yacoby.
\newblock Demonstration of entanglement of electrostatically coupled
  singlet-triplet qubits.
\newblock \emph{Science}, 336:\penalty0 202, 2012.
\newblock \doi{10.1126/science.1217692}.

\bibitem[Barthel et~al.(2009)Barthel, Reilly, Marcus, Hanson, and
  Gossard]{Barthel2009}
C.~Barthel, D.~J. Reilly, C.~M. Marcus, M.~P. Hanson, and A.~C. Gossard.
\newblock Rapid single-shot measurement of a singlet-triplet qubit.
\newblock \emph{Phys. Rev. Lett.}, 103:\penalty0 160503, 2009.

\bibitem[Reilly et~al.(2007)Reilly, Marcus, Hanson, and Gossard]{Reilly2007}
D.~J. Reilly, C.~M. Marcus, M.~P. Hanson, and A.~C. Gossard.
\newblock Fast single-charge sensing with a {RF} quantum point contact.
\newblock \emph{Appl. Phys. Lett.}, 91:\penalty0 162101, 2007.

\bibitem[Foletti et~al.(2009)Foletti, Bluhm, Mahalu, Umansky, and
  Yacoby]{Foletti2009}
S.~Foletti, H.~Bluhm, D.~Mahalu, V.~Umansky, and A.~Yacoby.
\newblock Universal quantum control in two-electron spin quantum bits using
  dynamic nuclear polarization.
\newblock \emph{Nature Physics}, 5:\penalty0 903, 2009.

\bibitem[Bluhm et~al.(2011)Bluhm, Foletti, Neder, Rudner, Mahalu, Umansky, and
  Yacoby]{Bluhm2011}
H.~Bluhm, S.~Foletti, I.~Neder, M.~S. Rudner, D.~Mahalu, V.~Umansky, and
  A.~Yacoby.
\newblock Dephasing time of {G}a{A}s electron-spin qubits coupled to a nuclear
  bath exceeding 200 $\mu$s.
\newblock \emph{Nature Physics}, 7:\penalty0 109, 2011.
\newblock \doi{DOI: 10.1038/NPHYS1856}.

\bibitem[Coish and Loss(2004)]{Coish2004}
W.~A. Coish and D.~Loss.
\newblock Hyperfine interaction in a quantum dot: Non-markovian electron spin
  dynamics.
\newblock \emph{Phys. Rev. B}, 70:\penalty0 195340, 2004.

\bibitem[Klauser et~al.(2006)Klauser, Coish, and Loss]{Klauser2006}
D.~Klauser, W.~A. Coish, and D.~Loss.
\newblock Nuclear spin state narrowing via gate-controlled {R}abi oscillations
  in a double quantum dot.
\newblock \emph{Phys. Rev. B}, 73:\penalty0 205302, 2006.

\bibitem[Ferrie et~al.(2013)Ferrie, Granade, and Cory]{Ferrie2013}
C.~Ferrie, C.~E. Granade, and D.~G. Cory.
\newblock How to best sample a periodic probability distribution, or on the
  accuracy of hamiltonian finding strategies.
\newblock \emph{Quantum Information Processing}, 12:\penalty0 611, 2013.

\bibitem[Schroer and Petta(2008)]{Schroer2008}
M.~D. Schroer and J.~R. Petta.
\newblock Quantum dots: time to get the nukes out.
\newblock \emph{Nature Physics.}, 4:\penalty0 1745, 2008.
\newblock \doi{10.1038/nphys1007}.

\end{thebibliography}
\end{document}


\title{Supplemental Material for{} ``Suppressing qubit dephasing using real-time Hamiltonian estimation''}

\date{\today}
\author{M. D. Shulman}
\author{S. P. Harvey}
\author{J. M. Nichol}
\affiliation{Department of Physics, Harvard University, Cambridge, MA,
  02138, USA}

\author{S. D. Bartlett}
\author{A. C. Doherty}
\affiliation{Centre for Engineered Quantum Systems, School of Physics,
  The University of Sydney, Sydney, NSW 2006, Australia}

\author{V. Umansky}
\affiliation{Braun Center for Submicron Research, Department of Condensed
  Matter Physics, Weizmann Institute of Science, Rehovot 76100 Israel}

\author{A. Yacoby}
\affiliation{Department of Physics, Harvard University, Cambridge, MA,
  02138, USA}
\email[email:]{yacoby@physics.harvard.edu}

\maketitle

\section{Singleshot Sensor Response}
\begin{figure}
\resizebox{\columnwidth}{!}{\includegraphics{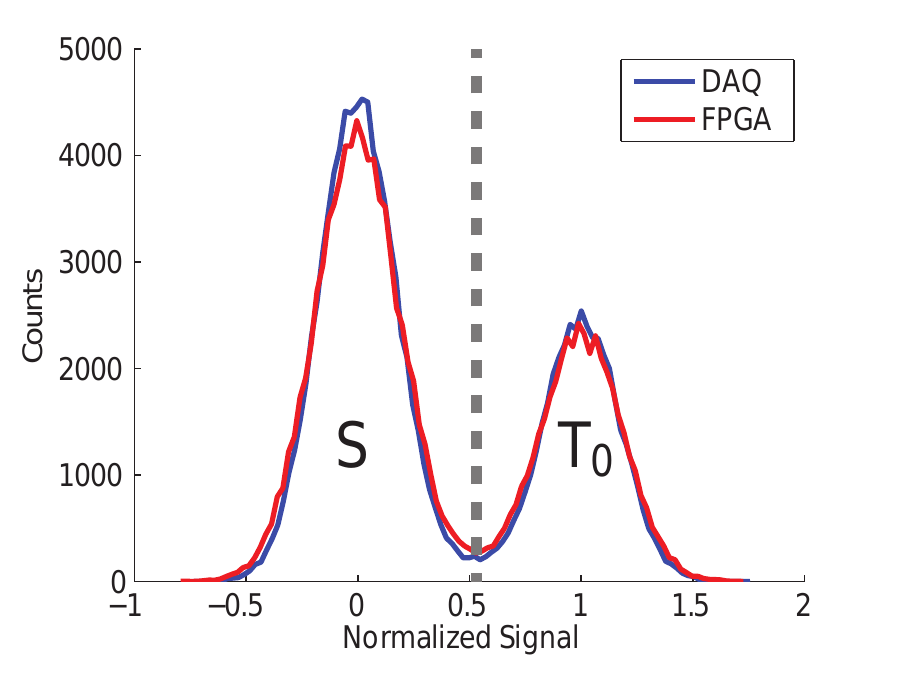}}
\caption{\textbf{a}. A histogram of values measured by the data
  acquisition card (DAQ) and the FPGA/CDS show nearly identical double
  peaked structures, indicating that they are capable of consistent
  singleshot readout. The dashed line is chosen as the threshold for
  estimating \dbz{} with the FPGA.
\label{rdout}
}
\end{figure}

In order to effectively sample \dbz{} oscillations without having to
measure each evolution time $t_k$ more than once, we rely on high
fidelity readout, which is based on standard RF-reflectometry
techniques\textsuperscript{16,17}. The readout fidelity is routinely
better than 0.98. Though the Bayesian estimate of \dbz{} has
parameters to account for readout error (see below), it nevertheless
requires that this error be small. Moreover, in order to effectively
process and compare data with both the FPGA and with software
rescaling, we must achieve high fidelity readout with both the data
acquisition card (DAQ) and with the FPGA. \figref{rdout}{a} shows
histograms of all of the measured values. The double-peaked structure
indicates that, indeed, high fidelity readout is achieved with both
the DAQ and the FPGA. The difference in the heights of the two peaks
is caused by residual exchange ($J$) during evolution, which causes
the axis of evolution around the Bloch sphere to be non-orthogonal to
the initial state (see section \ref{Bayes_section}). For the Bayesian
estimate, which requires discretized data ($r_{k}=\pm{}1$), we choose
a threshold corresponding to the minimum between the peaks for the
adaptive control on the FPGA.

\section{FPGA and experimental apparatus}
The reflected readout drive signal returns to room temperature through
a cryogenic circulator and amplifier at 4K. The signal is amplified
again at room temperature before being demodulated to DC. This DC
signal is split and sent to a digitizing card (AlazarTech 660) in a
computer and a home built correlated double sampler (CDS). The CDS
digitizes the signal and performs a local reference subtraction to
reject low frequency noise. The resulting 16 bit signal is converted
to a low voltage digital signal and sent to the FPGA for
processing. The FPGA is a National Instruments model PXI-7841R and is
clocked at 40MHz to maximize processing speed. The probability
$P(\dbz{}|m_{k})$ is computed for 256 consecutive frequencies in the
estimation bandwidth, $\mathcal{B}$, in two parallel processes on the
FPGA to decrease calculation time. Since $\mathcal{B}\approx{}$40MHz
is larger than the residual fluctuations of \dbz{}, we increase the
frequency resolution by computing the Bayesian estimate of \dbz{} for
the the middle 256 frequencies inside of $\mathcal{B}$. For these
parameters, the minimum calculation time is $3.7\mu{}s$ for a single
$t_k$. The probability distributions are stored and updated as
single-precision floating-point numbers, since we find that
single-precision improves the accuracy of the estimator over
fixed-point numbers.

\begin{figure}
\resizebox{\columnwidth}{!}{\includegraphics{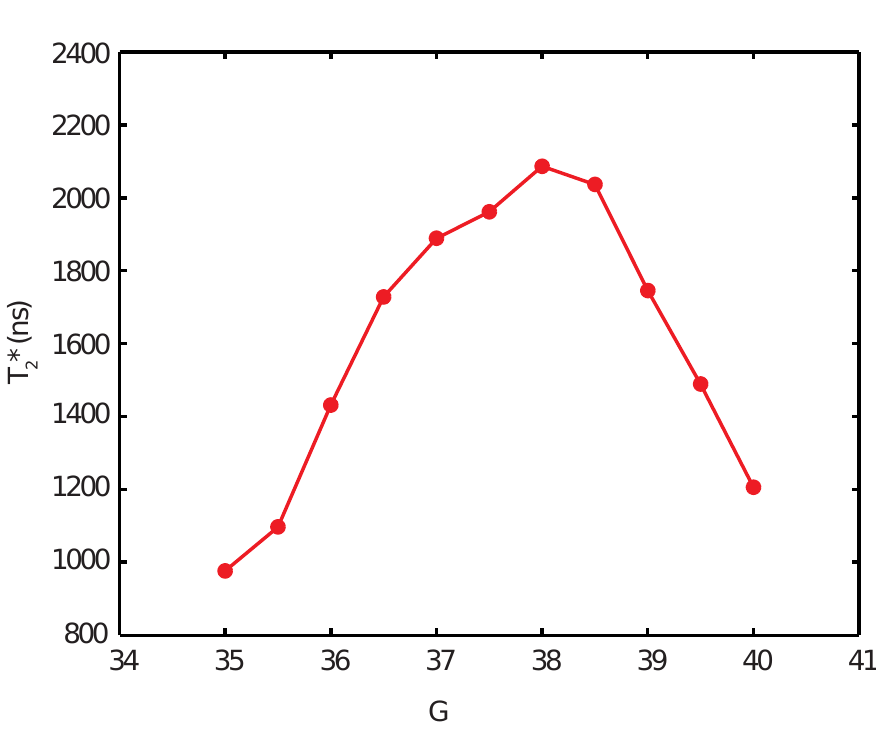}}
\caption{\textbf{a}. $T_{2}^{*}$ changes with the gain, $G$,
  converting a frequency index into a control voltage for the
  VCO. This allows for the optimal gain to be found.
\label{gain}
}
\end{figure}

After estimating \dbz{}, the FPGA returns the index (an integer
between 1 and 256) of the most probable frequency, which must be
converted to a voltage to control the VCO. To do so, we apply a linear
transformation to the index, $V=G\times{}index+O$, where the $O$
controls the detuning of the driving frequency. We tune the $G$ to
maximize $T_{2}^{*}$ using adaptive control (\figref{gain}{a}).

\section{Bayesian Estimate}
\label{Bayes_section}
We wish to calculate the probability that the nuclear magnetic field
gradient has a certain value, \dbz{}, given a particular measurement
record comprising $N$ measurements. We follow the technique in
Sergeevich \textit{et. al.}\textsuperscript{10} with slight
modifications. Writing the outcome of the $k^{\mathrm{th}}$
measurement as $m_{k}$, we write this probability distribution as
\begin{equation}
P\left(\dbz{}|m_{N},m_{N-1},...m_{1}\right).
\end{equation}
To arrive at an expression for this distribution, we will write down a
model for the dynamics of the system,
i.e. $P\left(m_{N},m_{N-1},...m_{1}|\dbz{}\right)$. Using Bayes' rule
we can relate the two equations as
\begin{eqnarray}
P\left(\dbz{}|m_{N},m_{N-1},...m_{1}\right)\cdot{}P\left(m_{N},m_{N-1},...m_{1}\right)\\=P\left(m_{N},m_{N-1},...m_{1}|\dbz{}\right)\cdot{}P\left(\dbz{}\right). \label{BayesRule}
\end{eqnarray}
First, we seek a model that can quantify
$P\left(m_{N},m_{N-1},...m_{1}|\dbz{}\right)$ that accounts for
realistic errors in the system, namely measurement error, imperfect
state preparation, and error in the axis of rotation around the Bloch
sphere. For simplicity, we begin with a model that accounts only for
measurment error. Denoting the error associated with measuring a
\ket{S} (\ket{T_0}) as $\eta{}_{S}$ ($\eta{}_{T}$), we write
\begin{eqnarray}
\nonumber P\left(S|\dbz{}\right)=\\
 (1-\eta_{S})\cos{}^{2}\left(2\pi{}\dbz{}t_{k}/2\right)+\eta_{T}\sin^{2}\left(2\pi{}\dbz{}t_{k}/2\right)\\ 
\nonumber P\left(T_{0}|\dbz{}\right)=\\
(1-\eta_{T})\sin{}^{2}\left(2\pi{}\dbz{}t_{k}/2\right)+\eta_{S}\cos{}^{2}\left(2\pi{}\dbz{}t_{k}/2\right)
\end{eqnarray}
We combine these two equations and write
\begin{equation}
P\left(m_{k}|\dbz{}\right) = \frac{1}{2}\left[1+r_{k}\left(\alpha +
  \beta{}\cos{}(2\pi{}\dbz{}t_{k})\right)\right]
\end{equation}
where $r_{k}$=1 (-1) for $m_{k}=\ket{S}$(\ket{T_{0}}) and $\alpha$ and
$\beta$ are given by
\begin{equation}
\alpha = \left(\eta_{T}-\eta_{S}\right),\; \; \beta{} =
\left(1-\eta_{S}-\eta_{T}\right).
\end{equation}
Next, we generalize the model to include the effects of imperfect
state preparation, and the presence of nonzero $J$ during evolution,
which renders the initial state non-orthogonal to the axis of rotation
around the Bloch sphere (see above). We assume that the angle of
rotation around the Bloch sphere lies somewhere in the $x$-$z$ plane
and makes an angle $\theta$ with the $z$-axis. We define
$\delta=\cos{}^{2}(\theta)$. Next, we include imperfect state
preparation by writing the density matrix
$\rho_{init}=(1-\epsilon{})\ket{S}\bra{S})+\epsilon{}\ket{T_{0}}\bra{T_{0}}$. With
this in hand, we can write down the model
\begin{eqnarray}
\nonumber
P\left(S|\dbz{}\right)=\eta_{T}+\frac{1}{2}(1-\eta_{S}-\eta_{T})
\times\\ 
\left\{1+(1-2\epsilon{})\left[\delta{}+(1-\delta{})\cos{}\left(2\pi{}\dbz{}t_{k}\right)\right]\right\},
\end{eqnarray}
\begin{eqnarray}
\nonumber
P\left(T_{0}|\dbz{}\right)=\eta_{S}+\frac{1}{2}(1-\eta_{S}-\eta_{T})
\times{}\\ 
\left\{1-(1-2\epsilon{})\left[\delta{}+(1-\delta{})\cos{}\left(2\pi{}\dbz{}t_{k}\right)\right]\right\}.
\end{eqnarray}
Using the same notation for $r_{k}$=1 (-1) for
$m_{k}=\ket{S}$(\ket{T_{0}}), we rewrite this in one equation as
\begin{equation}
P\left(m_{k}|\dbz{}\right) = \frac{1}{2}\left[1+r_{k}\left(\alpha +
  \beta{}\cos{}(2\pi{}\dbz{}t_{k})\right)\right], \label{model}
\end{equation}
where we now have 
\begin{eqnarray}
\alpha{}=\eta_{T}-\eta_{S}+(1-\eta_{S}-\eta_{T})(\delta{}-2\epsilon{}\delta{})\\
\beta{}=(1-\eta_{S}-\eta_{T})(1-\delta{})(1-2\epsilon{}).
\end{eqnarray}
We find the best performance for $\alpha=0.25$ and $\beta=0.67$, which
is consistent with known values for qubit errors.

We next turn our attention to implementing Bayes' rule to turn this
model into a probability distribution for \dbz{}. First, we assume
that all measurements are independent, allowing us to write
\begin{eqnarray}
P\left(\dbz{}|m_{N},m_{N-1},...m_{1}\right)&=& P\left(\dbz{}|m_{N}\right)\cdot{}P\left(\dbz{}|m_{N-1},...m_{1}\right) \nonumber\\
&=&\prod_{k=1}^{N}P\left(\dbz{}|m_{k}\right).
\end{eqnarray}
We next use Bayes rule (\ref{BayesRule}) and rewrite this equation as
\begin{equation}
P\left(\dbz{}|m_{N},m_{N-1},...m_{1}\right)=\prod_{k=1}^{N}P\left(m_{k}|\dbz{}\right)\frac{P(\dbz{})}{P(m_{k})}. 
\end{equation}
Using our model (\ref{model}) we can rewrite this as 
\begin{eqnarray}
P\left(\dbz{}|m_{N},m_{N-1},...m_{1}\right)=\\
\mathcal{N}P_{0}(\dbz{})\prod_{k=1}^{N}\left[1+r_{k}\left(\alpha{}+\beta{}\cos{}(2\pi{}\dbz{}t_{k})\right)\right], \label{postDist}
\end{eqnarray}
where $\mathcal{N}$ is a normalization constant, and $P_{0}(\dbz{})$
is a prior distribution for \dbz{} which we take to be a constant over
the estimation bandwidth, and to which the estimator is empirically
insensitive. With this formula, it is simple to see that the posterior
distribution for \dbz{} can be updated in real time with each
successive measurement. After the $N^{\mathrm{th}}$ measurement, we
choose the value for \dbz{} which maximizes the posterior distribution
(\ref{postDist}).

\section{Simulation with diffusion}

We simulate the performance of our software scaling and hardware
(FPGA) estimates of \dbz{} using the measured value of the diffusion
rate. We assume that \dbz{} obeys a random walk, but assume that
during a single evolution time $t_{k}$, \dbz{} is static. This
assumption is valid $\sqrt{t_{N}\mathcal{D}}T_{2}^{*}\ll{}1$, where
$\mathcal{D}$ is the diffusion rate of \dbz{}. For an estimation of
\dbz{} with $N$ different measurements, we generate a random walk of
$N$ different values for \dbz{} (using the measured diffusion),
simulate the outcome of each measurement, and compute the Bayesian
estimate of \dbz{} using the simulated outcomes. By repeating this
procedure 4096 times, and using the mean squared error,
$\mathrm{MSE}=\langle\left({}\dbz{}-\dbz{}^{estimated}\right)^{2}\rangle{}$
as a metric for performance, we can find the optimal number of
measurements to perform. To include the entire error budget of the
FPGA apparatus, we add to this MSE the error from the phase noise of
the VCO, the measured voltage noise on the analog output controlling
the VCO, and the diffusion of \dbz{} during the ``operation'' period
of the experiment.

\section{Software Post Processing}
\begin{figure}
\resizebox{\columnwidth}{!}{\includegraphics{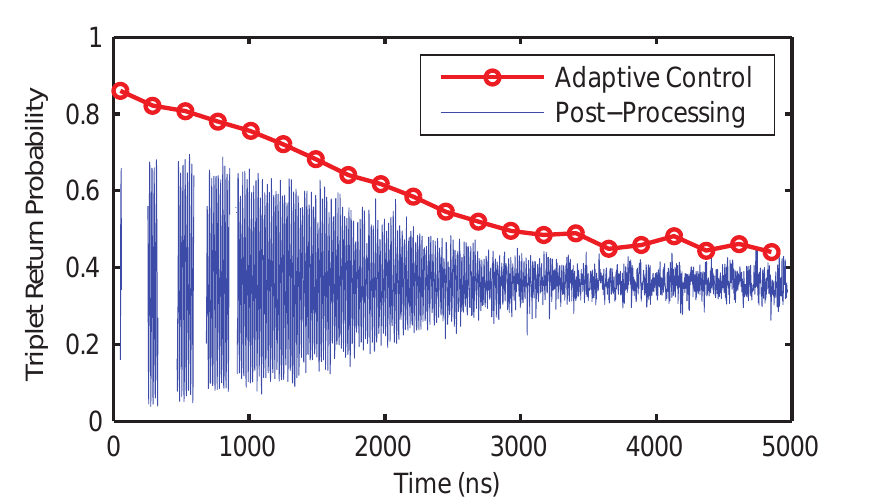}}
\caption{\textbf{a}. When using the same estimation sequence,
  post-processed oscillations (blue) and data taken using adaptive
  control (red) show the same decay, indicating similar performance
  of the estimation. The post-processing technique allows us to
  explore estimation sequences that are too fast for the FPGA.
\label{soft}
}
\end{figure}

The estimate of \dbz{} can be independently verified using software
analysis. In this experiment, we use the same method to estimate
\dbz{} as in the adaptive control experiment, but in the operation
segment perform oscillations around \dbz{} for verification. We choose
$m$ different evolution times and measure each $n$ times for a total
of $N_{tot}=m \times n$ measurements of \dbz. In the $i^{\mathrm{th}}$
experiment ($i = 1,2,\dots{}N_{tot}$), we evolve for a time
$t_{evo,i}$, accumulating phase $\phi_{i} = \dbz{}_{,i} t_{evo,i}
$. Because we make a precise measurement of \dbz{} at the start of
each experiment, we can employ it to rescale the time, $t_{evo,i}$, so
that the phase accumulated for a given time is constant using the
equation,

$$ \tau_i \equiv t_{evo,i} \frac{\dbz{}_{,i}}{\langle \dbz \rangle}$$
This sets $ \phi_i (\tau_i) = \langle \dbz \rangle \tau_i $, with
residual error arising from inaccuracy in the estimate of
$\dbz{}_{,i}$. The data are then sorted by $\tau$, and points of
similar $\tau$ are averaged using a Gaussian window with
$\sigma_{\tau} = 0.5 ~\mathrm{ns} \ll T \approx 16 ~\mathrm{ns}$,
where T is the period of the oscillations.

To compare post-processing with adaptive control, we first perform the
same estimation sequence for both software post-processing and
adaptive control, with a $250~\mathrm{kHz}$ repetition rate, $t_{samp}=
12 ~\mathrm{ns}$ and $N=$120, followed by an operation sequence of 30
measurements. We find $ T_2^* = 2148 \pm 30 ~\mathrm{ns}$ with
software and $T_2^* = 2066 ~\mathrm{ns}$ with adaptive control,
showing good agreement between the two approaches
(\figref{soft}{a}). 

For the software post-processing, we can reduce the amount of
diffusion that occurs during the operation sequence by performing only
one verification measurement following the same estimation sequence,
enhancing $T_2^*$, to $2580 \pm 40 ~\mathrm{ns}$. For the software
rescaling in Fig. 4d, the 109 estimations were performed in $ 225
~\mathrm{\mu s}$ instead of the $ 440 ~\mathrm{\mu s}$ used by the
FPGA, yielding $T_2^* = 2840 \pm 30 ~\mathrm{ns}$. This is likely
limited by diffusion and the precision of the estimator with $N$=109.